\documentclass[a4paper,tightenlines,preprint,prd,nofootinbib,floatfix,superscriptaddress]{revtex4-1}
\usepackage[T1]{fontenc}
\usepackage{amsmath,amssymb,amsfonts}
\usepackage{graphicx}
\usepackage{url,hyperref}
\usepackage[figtopcap]{subfigure}
\newcommand{\pwr}{p_\text{power}}
\newcommand{\re}[1]{\Re\left[ #1 \right]}
\newcommand{\im}[1]{\Im\left[ #1 \right]}

\providecommand{\href}[2]{#2}
\newcommand{\arxiv}[1]{arXiv:\href{http://arxiv.org/abs/#1}{{\tt #1}}}
\newcommand{\urlref}[1]{\href{#1}{\url{#1}}}
\newcommand{\SN}{\mathcal{N}}

\newcommand{\Omegac}[1]{\Omega^{(c)}_{#1}}
\newcommand{\Omegaf}[1]{\Omega^{(f)}_{#1}}
\newcommand{\Omegaa}[1]{\Omega^{(a)}_{#1}}

\newcommand{\xar}[1]{x^{(#1)}_\text{ar}}
\begin{document}
\title{Acceleration of the Arnoldi method and real eigenvalues of the
non-Hermitian Wilson-Dirac operator}
\author{Georg Bergner}
\email{g.bergner@uni-muenster.de}
\affiliation{Institut f\"ur Theoretische Physik, Westf\"alische
Wilhelms-Universit\"at M\"unster,
Wilhelm-Klemm-Str.\ 9, D-48149 M\"unster, Germany}
\author{Ja\"ir\ Wuilloud}
\email{wuilloud@gmail.com}
\affiliation{Albert Einstein Institute for Fundamental Physics\\
Institute for Theoretical Physics \\
University of Bern, Sidlerstr.\ 5, CH-3012 Bern, Switzerland}     
\date{\today}
\begin{abstract}
In this paper, we present a method for the computation of  the low-lying real
eigenvalues 
of the Wilson-Dirac operator based on the Arnoldi algorithm.
These eigenvalues contain information about several observables. 
We used them to calculate the sign of the fermion determinant in one-flavor QCD
and the sign of the Pfaffian in $\SN=1$ super Yang-Mills theory.
The method is based on polynomial transformations of the Wilson-Dirac operator,
leading to
considerable improvements of the computation of eigenvalues.
We introduce an iterative procedure for the construction of the polynomials and
demonstrate the improvement in the efficiency of the computation.
In general, the method can be applied to operators
with a symmetric and bounded eigenspectrum.
\end{abstract}
\pacs{XXX}
\maketitle
\section{Introduction}
The Wilson-Dirac operator $D_W$, which is used in many recent lattice
simulations to represent the fermionic part of the discretized action, 
has the following form
\begin{equation}
 (D_W)_{n,\alpha;m,\beta}=\delta_{n,m}\delta_{\alpha,\beta}-\kappa\sum_{\mu=1}^{4}
\left[(1-\gamma_\mu)_{\alpha,\beta}U_\mu(n)\delta_{n+\mu,m}+(1+\gamma_\mu)_{\alpha,
\beta}U^\dag_\mu(n-\mu)\delta_{n-\mu,m}\right]\, .
\end{equation}
Here $n,m$ denote points in a four-dimensional hypercubical space-time lattice, $\alpha, \beta$
are Dirac indices, $\mu=1,2,3,4$ labels the positive directions and $\gamma_{\mu}$
are the Dirac matrices. The hopping parameter $\kappa$ is related to the bare
fermion mass; in particular $\kappa$ increases for decreasing fermion masses.
The link variables $U_{\mu}(n)$ are associated with the links
connecting neighboring lattice points and represent the gauge field.
In our investigations 
\cite{Farchioni:2007dw,Demmouche:2010sf}
the gauge field was in the fundamental representation of SU(3) for QCD with one quark flavor 
(one-flavor QCD) and in the adjoint representation of SU(2) (real $3\times3$ matrices) 
for supersymmetric Yang-Mills theory. The method presented here is,
however, not restricted to a specific gauge group and can be applied also to
other fermion operators. 

In the free theory, the eigenspectrum of $D_W$ can be decomposed into a physical
branch, consisting of the smallest eigenvalues, and the doublers, which become
irrelevant in the continuum limit \cite{Creutz:2005rd}.
Such a clear distinction of relevant and irrelevant parts is not possible in the
interacting case. 
However, the lowest part of the spectrum still contains the most important
information.
The low eigenvalue part  plays a crucial role  in spectral decompositions of the
fermionic observables \cite{Neff:2001zr}, and the lowest eigenmodes allow for an
acceleration of the inversion by deflation \cite{Wilcox}.

For several investigations, the Hermitian operator $\gamma_5 D_W$ can be used
instead of $D_W$.
The corresponding eigenvalue problem can also be solved with other iterative
methods, but for the non-normal operator $D_W$ the (restarted) Arnoldi algorithm
\cite{ARPACK} seems to be the optimal choice.\footnote{See, e.\,g.,
\cite{Hip:2001mh} for a detailed discussion of the effects of the non-normality.
}

The importance of the lowest eigenmodes of $D_W$ has been the subject of several
recent investigations, e.\,g.\ in \cite{Bruckmann,Synatschke}.
Furthermore, their implications on the topology of gauge fields has been studied (e.\,g.\ in
\cite{Topology1,Topology2}),
even though $D_W$ does not allow for a realization of chiral symmetry on the
lattice.

In several cases, numerical simulations of field theories with dynamical fermions require  a
reweighting of the observables with the sign of the determinant of $D_W$ 
or of its Pfaffian. This sign can be obtained from the number of negative real eigenmodes
\cite{Farchioni:2007dw,jaja}.
The computation of the reweighting for one-flavor QCD and $\SN=1$ super Yang-Mills theory was the
main purpose of our investigations of the spectrum of $D_W$.

On small lattices, the complete set of eigenvalues is accessible (see e.\,g.\
Fig.~\ref{fig:completespec1}).
In a more realistic setup, strategies focusing the computation on the relevant
small eigenvalues and
accelerating the convergence are required.
For lattice QCD, a polynomial approach focusing on the low eigenmodes of $D_W$
has been presented in \cite{Neff}.
Within a mathematical framework, other methods based on polynomial
transformations have been developed for the computation of a particular sector
of a general eigenspectrum \cite{saad,heuvelinearnoldifaber}.
We explain here a new strategy to obtain the lowest real eigenmodes of the
Wilson-Dirac operator and show its impact on the efficiency of the computation.
Our strategy allowed us to obtain the relevant part of the spectrum on lattices
up to a size of $32^3\times 64$ lattice points.

This paper is organized as follows. In the next section we explain the basic
idea of the polynomial transformations.
In Section \ref{sec:peeling} we present our specific strategy to obtain the necessary
polynomial. Section \ref{sec:results} contains some results and
Section \ref{sec:comparison}  a comparison with other methods. Further
mathematical explanations and some practical considerations can be found in
Section \ref{sec:tech}.
\section{Acceleration and focusing of the Arnoldi algorithm}
\label{sec:lattice formulation}
\begin{figure}[ht]
\begin{center}
        \includegraphics[width=14cm]{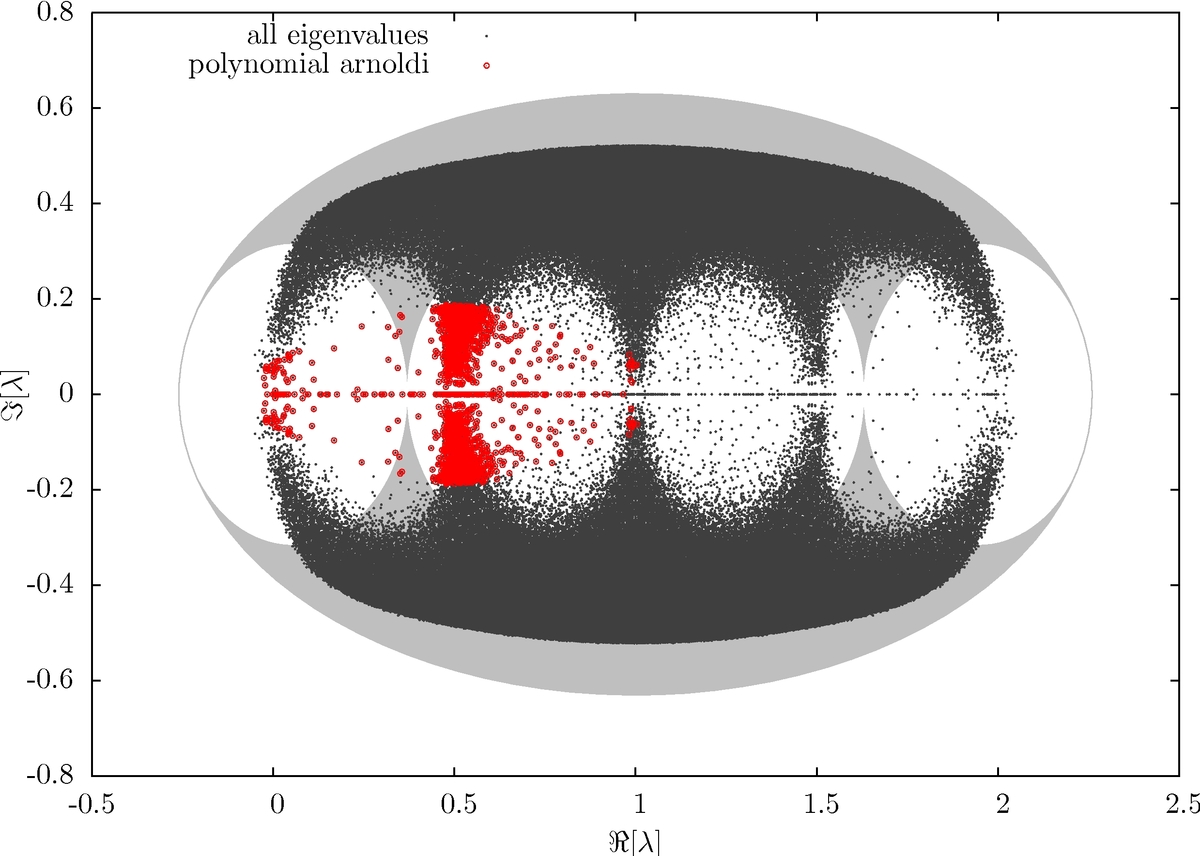}
		\caption{
        As an illustration of the method and the form of the spectrum to be
expected,
	this figure shows all eigenvalues of the Wilson-Dirac operator $D_W$
from $29$ independent thermalized configurations of a dynamical simulation of the
$\SN=1$ super-Yang-Mills theory \cite{Demmouche:2010sf} on a $6^3\times 8$
lattice ($\beta=1.6$; $\kappa=0.1575$). The red points
correspond to the eigenvalues computed with the peeling method described below.
The gray background shows the region of the eigenvalues in the free theory with
the same value of $\kappa$.
	}
	\label{fig:completespec1}
\end{center} \end{figure}
Let $\Omegac{D_W}$ be the region that contains all eigenvalues of the operator.
We are only interested in a subset of eigenvalues enclosed in a region denoted by
$\Omegaf{D_W}$. 
In our case, $\Omegaf{D_W}$ is chosen to be a prolate region surrounding
all real eigenvalues smaller (or larger) than a
certain value (e.~g.\ $\Omegaf{D_W}=\{\lambda_i| \,
|\im{\lambda_i}|<\varepsilon,\re{\lambda_i}\geq x_\text{min}\}$ with $\varepsilon$ small).

Fig.~\ref{fig:completespec1} shows that the spectrum of $D_W$ contains large
regions with a high eigenvalue density and a nonzero imaginary part.
For an efficient calculation of the real eigenvalues, it is crucial to exclude
these regions and focus the computation on the eigenvalues in $\Omegaf{D_W}$. 
The Arnoldi algorithm computes the eigenvalues starting from those with  largest
real part. It calculates the eigenvalues in the region $\Omegaa{D_W}=\{\lambda |
\re{\lambda}>\xar{D_W}\}$, where $\xar{D_W}$ depends on the parameters of the
algorithm and the eigenvalue distribution of $D_W$.

Hence, a direct computation does not focus efficiently enough on $\Omegaf{D_W}$
since a lot of unwanted eigenvalues are calculated.
However, an appropriate polynomial transformation $D_W \to p(D_W)$ leads to a
better overlap of $p(\Omegaf{D_W})$ with $\Omegaa{p(D_W)}$.
The computation gets focused on the relevant part and a smaller number of
unwanted eigenvalues is computed.
The eigenvalues of $D_W$ can be obtained from the eigenvalues or eigenvectors of
$p(D_W)$.

The second advantage of the polynomial transformation is an acceleration of the
Arnoldi computation. 
The computation of an eigenvalue $\lambda$ converges faster, if this eigenvalue 
is better separated from the rest of the spectrum (compared to some
average distance of the eigenvalues).
Therefore, a polynomial minimized on $\Omegac{}\setminus \{ \lambda\} $ (with
$p(\lambda_i)$ fixed) leads to an acceleration of the computation of $\lambda_i$
(for details cf.\ \cite{saad,Beattie:2005}). An analytic solution for the
absolute minimum on a general $\Omegac{}\setminus\{\lambda \}$ is not available, but Chebyshev
\cite{saad} and Faber polynomials \cite{heuvelinearnoldifaber} provide
approximate solutions of it.

Since the algorithm starts from a random initial vector, it can happen that some
eigenvalues within $\Omegaa{D_W}$ are not found in the Arnoldi iteration.
Especially, some in a set of closely lying or exactly degenerate
eigenvalues might be missing. 
This effect is considerably reduced by the polynomial transformation.

For an appropriate polynomial, the focusing effect and the acceleration
by far compensate the costs of the additional multiplications.
Eigenvalues in the original spectrum obtained with a polynomial transformation
are shown in Fig.~\ref{fig:completespec1}.
\section{The peeling transformation}
\label{sec:peeling}
In previous investigations of the lowest real eigenvalues of $D_W$ in lattice
QCD \cite{Neff}, a certain set of simple polynomials has been proposed.
It consists of power transformations of the form 
\begin{equation}
\pwr(D_W;n,\sigma)= \left( D_W + \sigma \mathrm{I} \right)^n , \, \text{with} \,
n \in \mathbf{N}, \, \sigma \in \mathbf{R}\, . \label{power_trafo}
\end{equation}
It has been shown to considerably improve the performance of Arnoldi algorithm.
\footnote{A choice of parameters $n$ and $\sigma$ is explained in
Sec.\ \ref{sec:tech}.}
The effect of this transformation on a test eigenspectrum is illustrated  in 
Fig.~\ref{fig:step1peel}.
The region of computed eigenvalues in the original spectrum gets a wedge like
shape. Hence, the computation is better focused on $\Omegaf{D_W}$.
However, at larger $n$ the focusing effect saturates.

Based on these observations, we propose here the ``peeling transformation'' as
an iteration of the power transformation. 
It consists in the following steps:
\begin{enumerate}
\item The starting point is a power transformation with an additional
renormalization factor $r_0 \in \mathrm{R}$,
   $ p_0(D_W;n_0, \sigma_0,r_0)=\pwr(D_W/ r_0;n_0,\sigma_0) $.
\item For the resulting eigenspectrum, a new power transformation is chosen for
a further focusing on $\Omegaf{D_W}$, \\ $
p_1(D_W;n_0,\sigma_0,r_0,n_1,\sigma_1,r_1)=\pwr\left( p_0(
D_W;n_0,\sigma_0,r_0)/r_1;n_1,\sigma_1)\right)$.
\item  This procedure is iterated until the polynomial $p_N$ is obtained. 
\end{enumerate}
The effect of the further iterations on a test eigenspectrum is shown in 
Fig.~\ref{fig:transformationsteps}.
Clearly, the eigenvalues in $\Omegaf{D_W}$ are made accessible by the
transformation, while $\Omegac{D_W}\setminus \Omegaf{D_W}$ is compressed in a
region close to the transformed zero.
The polynomial resulting from the iteration is
\begin{equation}
p_N(D_W;n_0,\sigma_0,r_0,\ldots,n_N,\sigma_N,r_N)= (\cdots( ( D_W/r_0 + \sigma_0
\mathrm{I})^{n_0}+\ldots)/r_N+\sigma_N)^{n_N}\, ,
\end{equation}
with the free parameter $n_i$, $\sigma_i$, and $r_i$.\footnote{The parameter
$r_i$ can be absorbed into a redefinition of the $\sigma_i$ and an overall
rescaling.}
The question of an optimal choice for these parameters depends on the form of
the eigenspectrum and is addressed
in Sec.\ \ref{sec:tech}.
\begin{figure}
\addtocounter{figure}{1}
\setlength{\subfigcapskip}{-0.4cm}
\setlength{\subfigbottomskip}{0.5cm}
\newlength{\pichg}
\setlength{\pichg}{4.55cm}
  \begin{center}
\subfigure[The initial test spectrum with equidistant
eigenvalues.]{\includegraphics[height=\pichg]{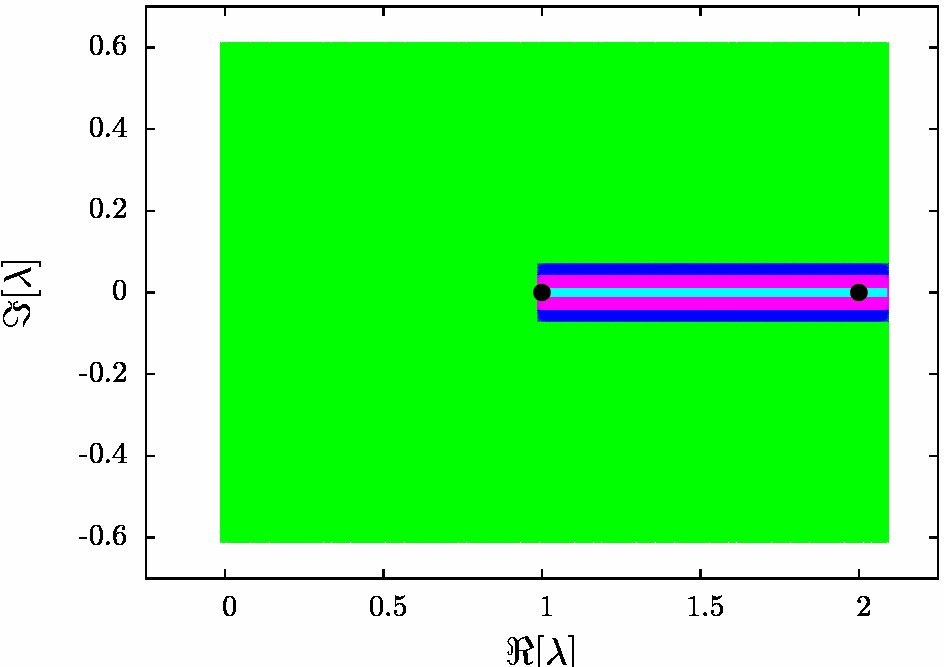} \includegraphics[height=\pichg]{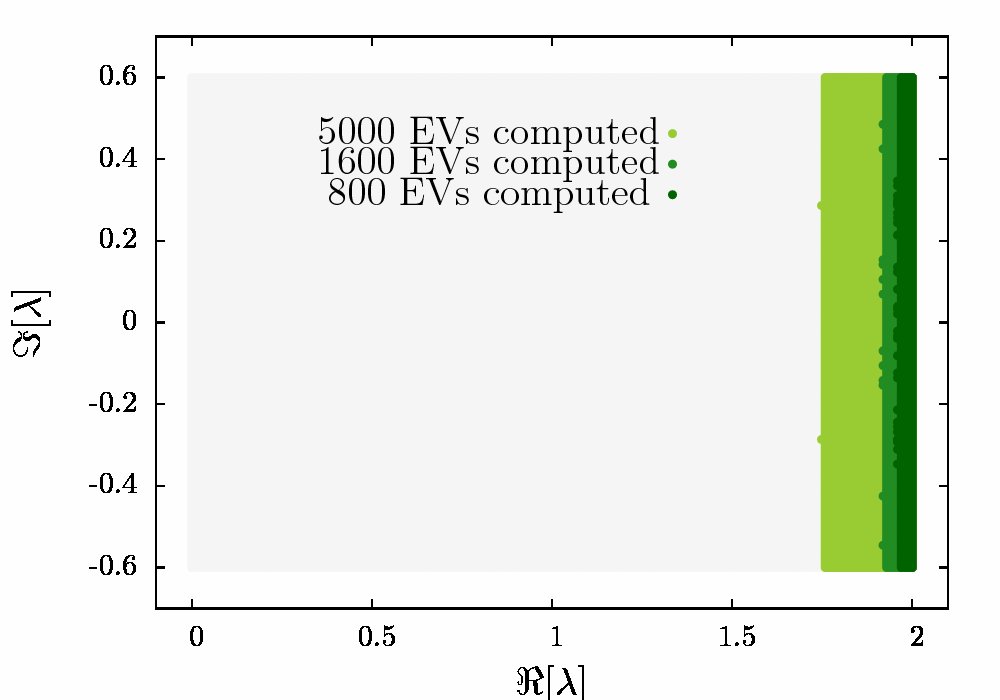}}
\subfigure[$p(x)=p_0(x;9,1,1)$]{ \includegraphics[height=\pichg]{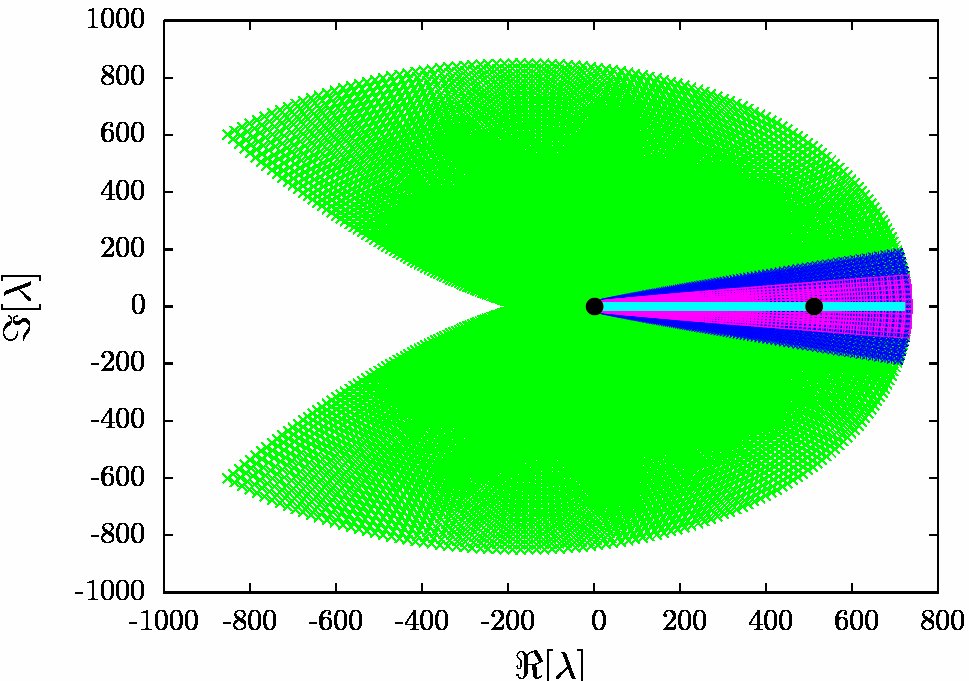} \includegraphics[height=\pichg]{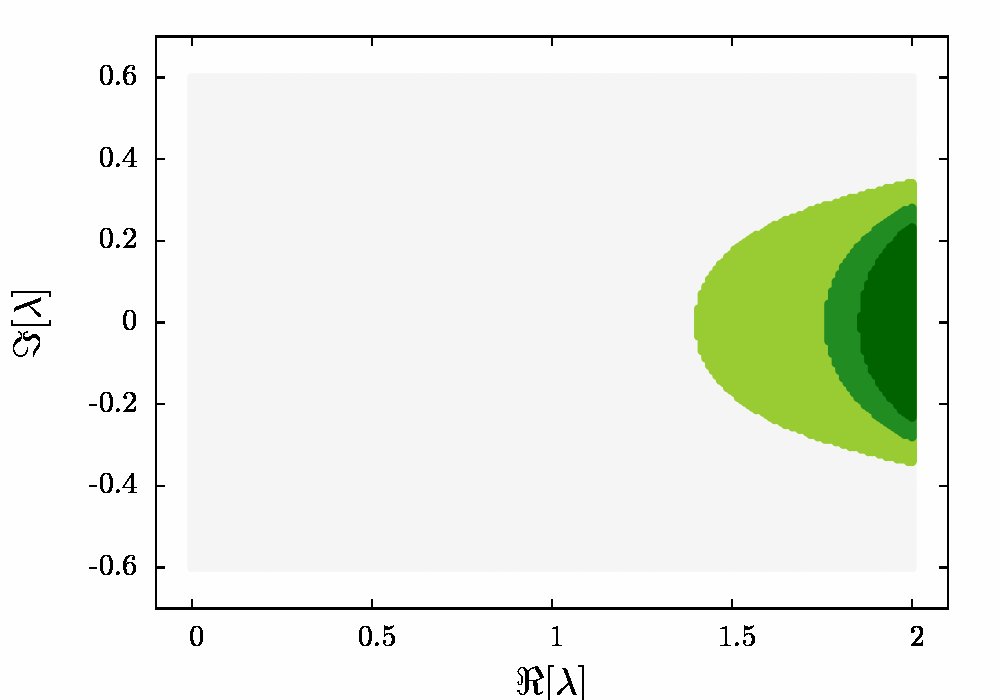}\label{fig:step1peel}}
\subfigure[$p(x)=p_1(x;9,1,1,6,2,500)$]{\includegraphics[height=\pichg]{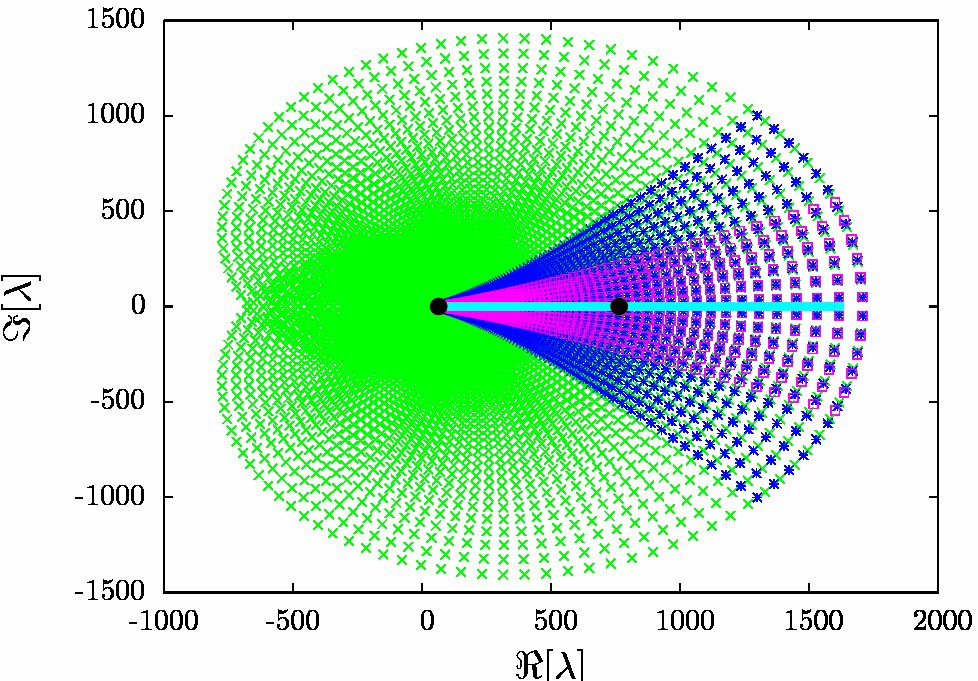} \includegraphics[height=\pichg]{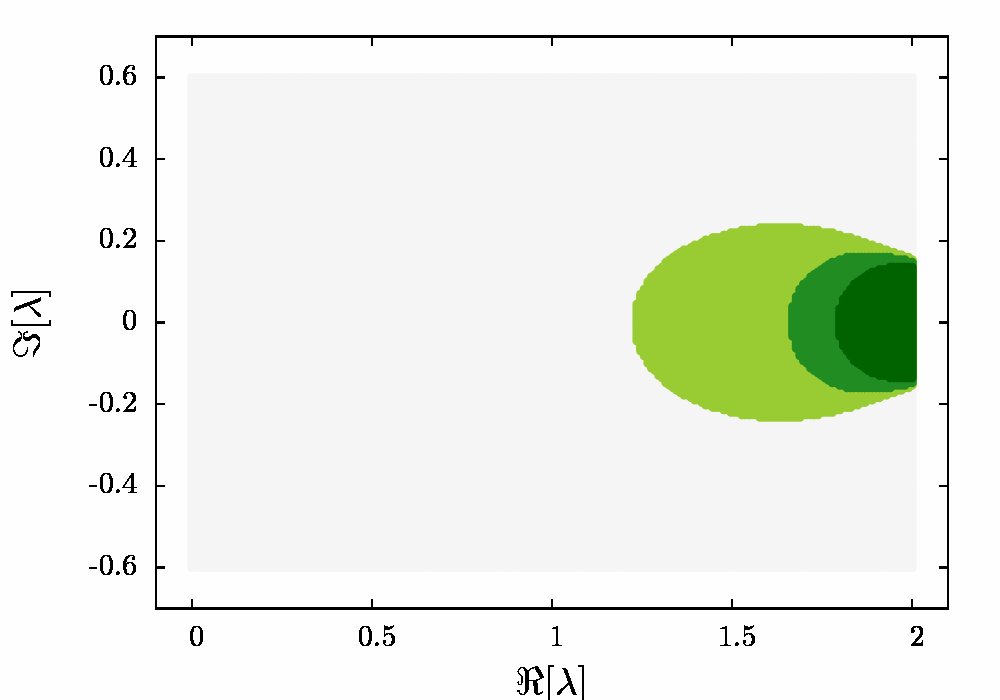}}
\subfigure[$p(x)=p_2(x;9,1,1,6,2,500,2,16,1000)$]{ \includegraphics[height=\pichg]{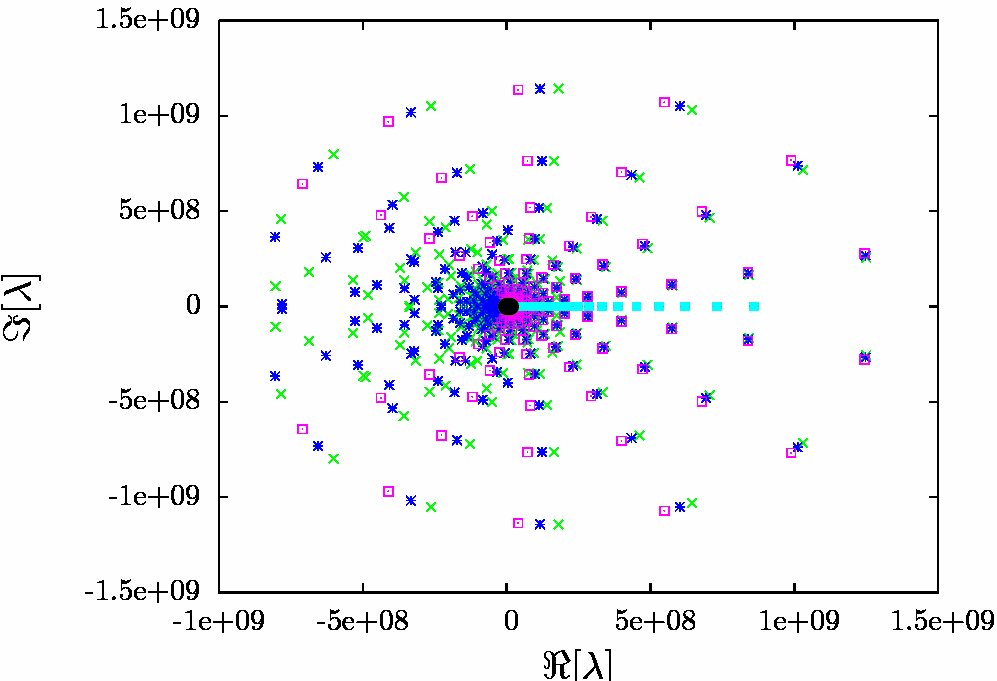} \includegraphics[height=\pichg]{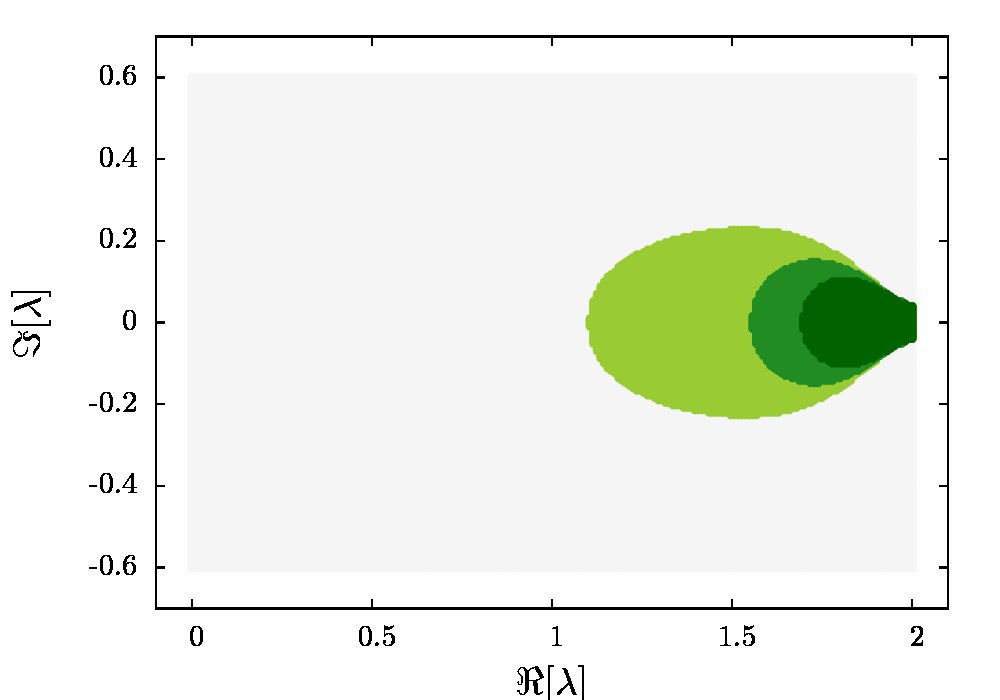}}
\addtocounter{figure}{-1}
\caption{
The figures on the left hand side show the result of the polynomial transformation 
applied to a test spectrum of equidistant eigenvalues that fill a rectangular region.
The small region colored in light blue corresponds to $\Omegaf{D_W}$.
The figures on the right display the parts in the original spectrum computed 
successively in the polynomial Arnoldi iterations.
In the first figures no polynomial is applied. The polynomials applied in the other 
figures are listed below each of them.
}\label{fig:transformationsteps}
\end{center}\end{figure}
\section{Real eigenvalues and determinant signs}
\label{sec:results}
\begin{figure}[ht]
\includegraphics[width=8cm]{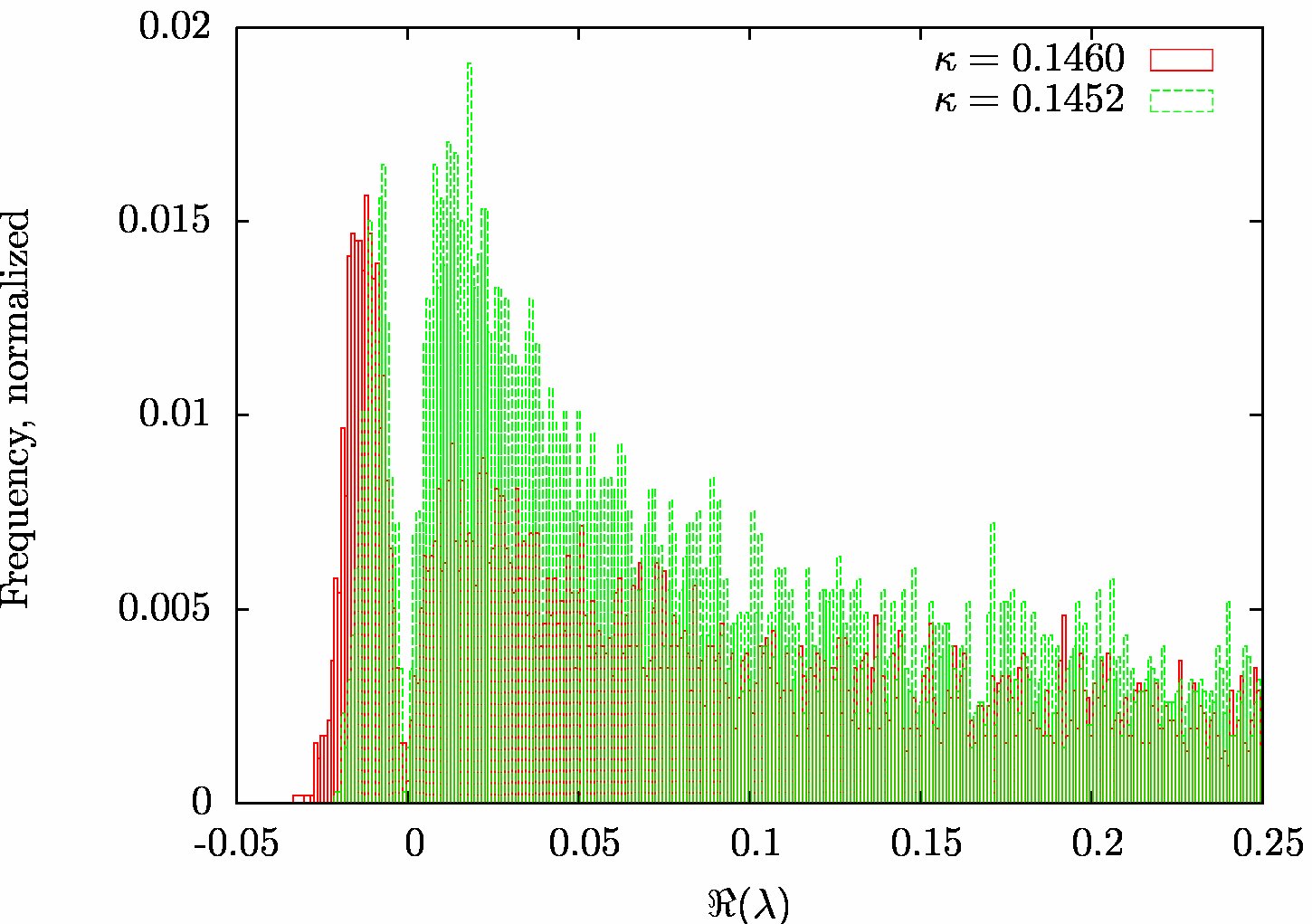}
\caption{Distributions of the lowest real eigenvalues of the even-odd
preconditioned Wilson-Dirac operator in simulations of one-flavor QCD.
  When $\kappa$ is increased, the distribution is shifted and a larger peak
in the negative sector is observed.}
\label{lowestEV8x16}
\end{figure}

One of the goals of our calculations were the determinant signs for numerical
simulations of one-flavor QCD and Pfaffian signs for the supersymmetric Yang-Mills theory.
In order to realize small pion or gluino masses, both theories were simulated
within a parameter regime where
very small and negative eigenvalues appear.
Except for the real modes, all eigenvalues of $D_W$ appear in complex conjugate pairs.
Thus the determinant and Pfaffian signs depend only on the real negative eigenvalues; in particular
\begin{equation}
\mathrm{det}
(D_W)= \prod_{ \lbrace \lambda \in \mathbf{C} \, \mid \, \im{\lambda} >0\rbrace
} \mid \lambda \mid^2 \prod_{ \im{\lambda} =0} \lambda,
\end{equation}
see \cite{Montvay:2001aj,Campos:1999du} for more details.
For one-flavor QCD, Fig.~\ref{lowestEV8x16} illustrates 
the distribution of the lowest eigenvalues for two different $\kappa$ values.

This method, based on a direct computation of the real negative eigenvalues of
$D_W$, turns
out to be more efficient than the previously considered ``eigenflow''
approach, where determination of the sign is based on the eigenvalues of $\gamma_5
D_W$. 
This Hermitian matrix allows to compute its eigenvalues by means of
simpler computational methods. 
However, in order to obtain the determinant (Pfaffian) sign the eigenvalues have to be
computed at several different $\kappa$ values
\cite{Campos:1999du,Montvay:2001aj}.\footnote{This method is similar to the
computation of all real eigenvalues in \cite{EigenFlow1}.}

Depending on the parameters of the simulation, in particular on the value of
the hopping parameter $\kappa$, we obtained up to $50\%$ negative
signs in the simulation of one-flavor QCD. With increasing $\kappa$,
approaching its critical value $\kappa_c$, the number of negative signs
increases.
For values of $\kappa$, which have been used for
measurements of the particle spectrum,
however, the number of negative signs was well below $10\%$.
\section{Comparison of the polynomial transformations}
\label{sec:comparison}
To demonstrate the importance of the different steps in the peeling method, we compare
the performance of several different peeling polynomials with power polynomials. The
computation time needed to obtain a number of wanted eigenvalues allows for a
simple 
and clear representation of this performance. In the present case,
$\Omegaf{D_W}$ is the region of all eigenvalues with an imaginary part whose absolute value is
smaller than $0.05$. The polynomials were constructed as described in Section
\ref{sec:tech}.

\begin{figure}[ht]
  \includegraphics[width=8.1cm]{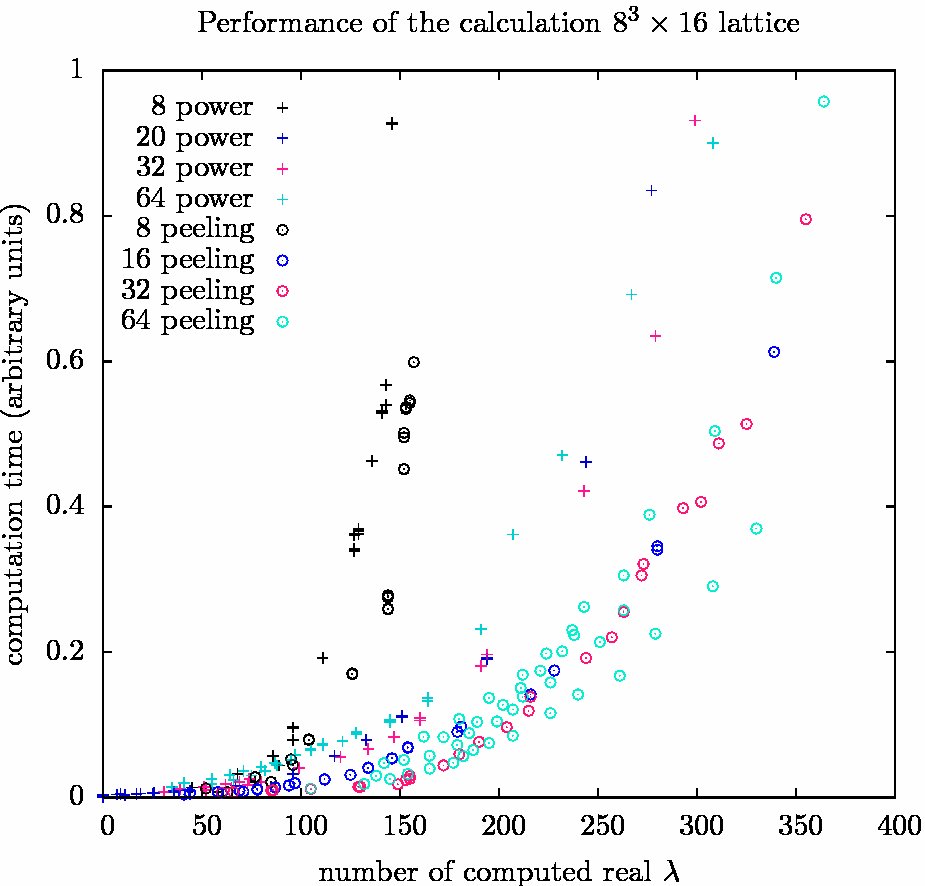}
  \includegraphics[width=8.1cm]{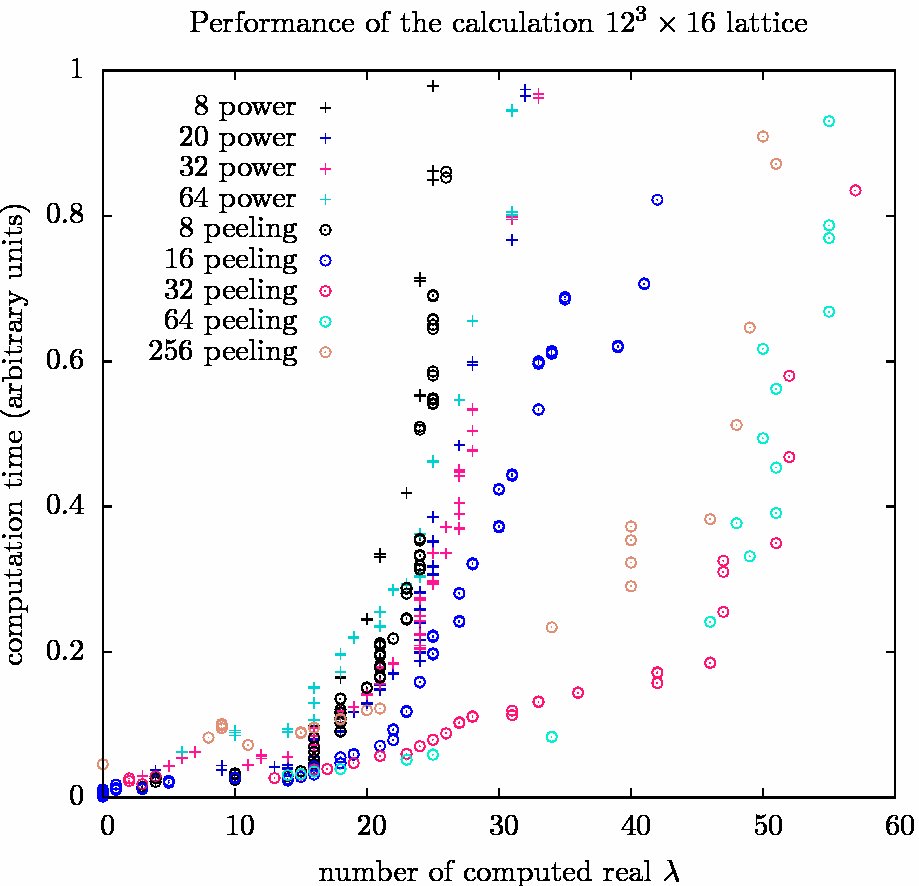}
\caption{
The performance of the different polynomials with different sizes of the
eigenspace and different maximal number of iterations.
The numbers in front of the polynomial name indicate the polynomial order.
The computations were done on $91$ / $11$ configurations of one-flavor QCD for
the $8^3\times 16$ / $12^3\times 16$ lattice.
All eigenvalues counted for the number of real $\lambda$ have an imaginary part
smaller than $0.05$.
The large variation for the order $64$ peeling polynomial on the small lattice
shows a dependence on the maximal number of iterations as in 
Fig.~\ref{fig:comparison2}.
}
\label{fig:comparison1}
\end{figure}

The eigenvalue computations depend on the number of computed eigenvalues, the
size of the available eigenspace, and the maximal number of iterations. We have
varied all of these parameters for the comparison of the polynomials. 
Fig.~\ref{fig:comparison1} shows the performance of different orders of the power
method compared to peeling polynomials of a similar order. Clearly the peeling
polynomials allow for a more efficient calculation of the eigenvalues in the
considered region. The improvement of the Arnoldi extraction by the polynomials
seems to be saturated at a certain order. The extra number of matrix vector
multiplications compensates the focusing and acceleration effect. 
This saturation happens at higher orders for the peeling polynomials than for
the power polynomials. 
In case of the peeling polynomials, the saturation depends on the eigenvalue
density, since for a larger lattice size it happens at a larger order.
At a smaller lattice size the eigenvalue density seems to be too low to profit
from the better focused calculation.
Eventually the performance is limited when the next region of a high eigenvalue 
density is reached. In the considered spectra these regions form a regular pattern 
similar to the free theory (cf.\ Fig.~\ref{fig:completespec1}). 
Therefore, the limiting high eigenvalue density can be attributed to the first doublers. 
A steep rise of computation time is visible at this point,
especially on larger lattices.

\begin{figure}[ht]
  \includegraphics[width=8.1cm]{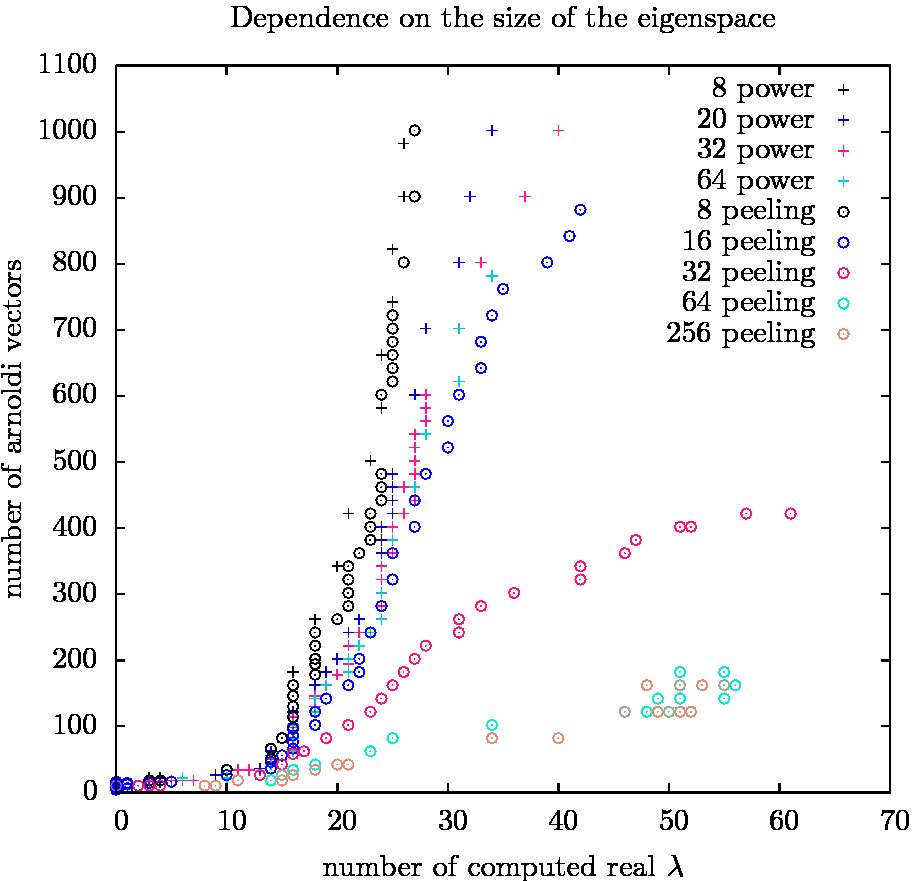}
  \includegraphics[width=8.1cm]{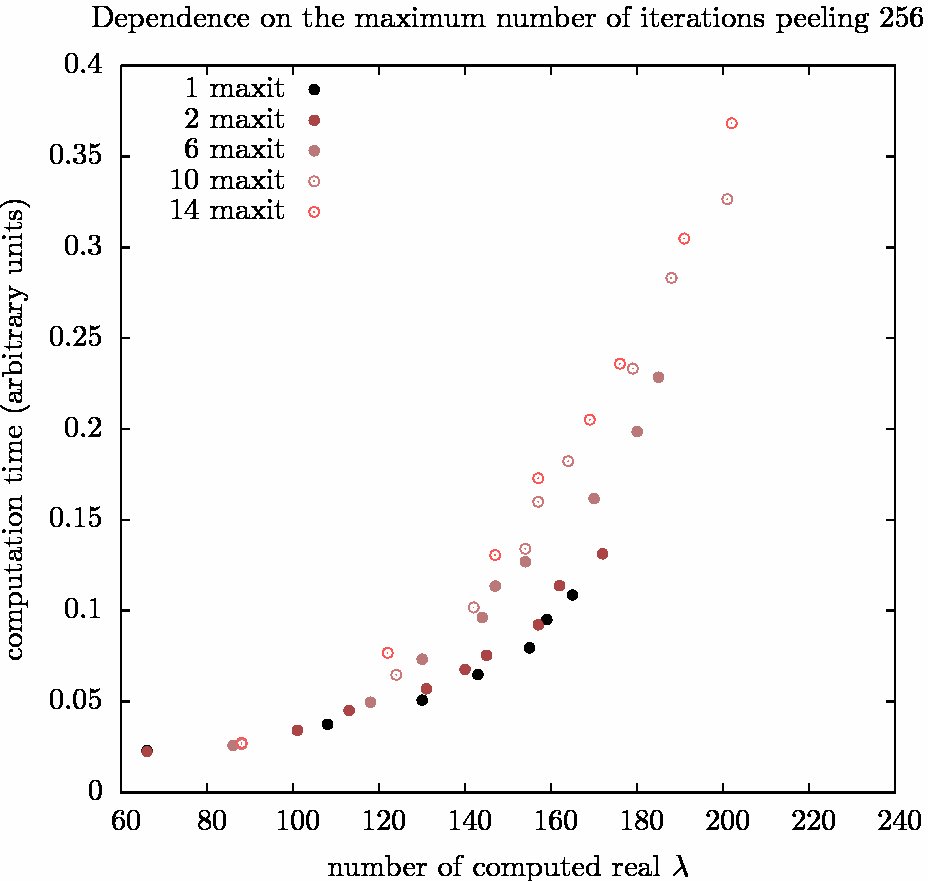}
       \caption{
        The figure on the left hand side illustrates the memory (number of
vectors) required for the computation of a number of wanted eigenvalues ($12^3\times 16$ lattice).
        The eigenspace size was set to twice the number of considered
eigenvalues plus two.
        On the right, we show the dependence on the maximal number of iterations
allowed in the Arnoldi algorithm for an order $256$ peeling polynomial ($8^3\times 16$ lattice).
        In this figure the number of considered eigenvalues is enlarged, but the
maximum number of iterations is kept at a small value.
        }
        \label{fig:comparison2}
 \end{figure}

Besides the time of the computation the required memory can be a limitation of
the eigenvalue computations. In that respect, the peeling approach exhibits
decisive advantages with respect to the power method. This is shown in the right
part of Fig.~\ref{fig:comparison2}, where the needed memory is represented by
the number of vectors used in the computation. With respect to this requirement
even quite large orders of the peeling polynomials can lead to an improvement.

Using larger orders of the peeling polynomials on smaller lattices one observes
that most of the time for the Arnoldi computation is spent on a certain set
of configurations with a low eigenvalue density inside $\Omegaf{D_W}$. To avoid
this effect one can imply a small limit on the maximal number of iterations such
that a smaller number of eigenvalues is extracted on these rather uninteresting
configurations.
The resulting improvement is shown in the right part of 
Fig.~\ref{fig:comparison2}. 
Nevertheless, one should be careful with the limit on the maximum number of
iterations. The chance of missing some eigenvalues is increased when the number
of iterations gets very low.
The Arnoldi algorithm requires a balance of the number of multiplications in the
polynomial and of the Arnoldi iterations.
\section{Technical details of the polynomial design}
\label{sec:tech}
The polynomials can be designed to improve the focusing on the largest or the
smallest real eigenvalues.
To simplify the notation, we assume that in the latter case the transformation
$D_W\rightarrow 2-D_W$ is applied, such that again the largest real eigenvalues
should be computed.\footnote{For $D_W$, this is a symmetry transformation and
the spectrum remains unchanged, but we applied the method also for other
operators, like the even-odd preconditioned Wilson-Dirac operator.}
The region of the wanted eigenvalues is hence $\Omegaf{D_W}=\{\lambda_i| \,
|\im{\lambda_i}|<\varepsilon,\re{\lambda_i}\geq x_\text{min}\}$, where
$\varepsilon>0$ is small and $x_\text{min}$ is the deepest point in the spectrum
considered in the computation. 
For simplicity, the normalization of the polynomial is chosen such that
$p(x_\text{min})=1$ in each step of the peeling transformation.
Thus, $\sigma_k$ is replaced using $\sigma_0=1-x_\text{min}/r_0$ or
$\sigma_k=1-1/r_k$ for $k>0$.

To understand the effect of one step of the peeling transformation (i.~e.\ a
power transformation), we represent the complex eigenvalues $\lambda_i$ by their
radius  and phase,
\begin{equation*}
\rho_i=\sqrt{(\re{\lambda_i}/r_0-\sigma_0)^2+(\im{\lambda_i}/r_0)^2} \text{ and
} \theta_i=\arctan\left[(\re{\lambda_i}-\sigma_0r_0)/\im{\lambda_i}\right]\, ,
\end{equation*}
after the shift and rescaling. 
The phase is mapped onto $n_0\theta_i$ and a fraction of the eigenvalues with a
nonzero imaginary part are hence ``rotated away'' from the real axis and out of
the region of the computed eigenvalues. This effect focuses the calculation on
the real eigenmodes. 
However, the eigenvalues with the largest $\theta$ can be ``rotated'' inside the
computed region. Let $\theta_\text{max}$ be the maximal phase of all $\lambda_i$
with  $\rho_i\geq 1$.
One way to avoid such an entering of the ``rotated'' eigenvalues is to apply the
restriction $n_0\theta_\text{max}< 3\pi/2$.

The focusing effect can be better controlled when it is visualized by a plot of
the contour $\re{p(\lambda)}=1$ in the complex plane.
The eigenvalues in the region of all $\lambda$ with $\re{p(\lambda)}\geq 1$,
i.~e.\ inside the contour, are computed by the algorithm when it reaches the
real eigenvalue $\lambda=x_\text{min}$. 
 There are $n_0$ of such regions, and the contours surrounding them tend for
$\rho\rightarrow \pm\infty$ to the lines $\theta=(2l+1)\pi/(2 n_0)$, with
$l=0,1,\ldots,n_0-1$.
The larger the number of eigenvalues in $\Omegaf{D_W}$ divided by the number of
eigenvalues in these regions, the better is the focusing of the polynomial.
The restriction to avoid an entering of the ``rotated'' points in the computed
region can now be made more precise:
the parameters are restricted such that only the region of one contour
surrounding $\Omegaf{D_W}$ has overlap with $\Omegac{D_W}$.
This region should be made as small as possible for the best focusing. 
Thus, for a given $n_0$, $r_0$ must be minimized as much as possible without the
appearance of a second contour inside $\Omegac{D_W}$.\footnote{The 
minimization of $r_0$ leads to a maximal $\theta$ and hence to a maximal effect of 
the ``rotation''.}

\begin{figure}[ht]
\begin{center}
\includegraphics[width=8.1cm]{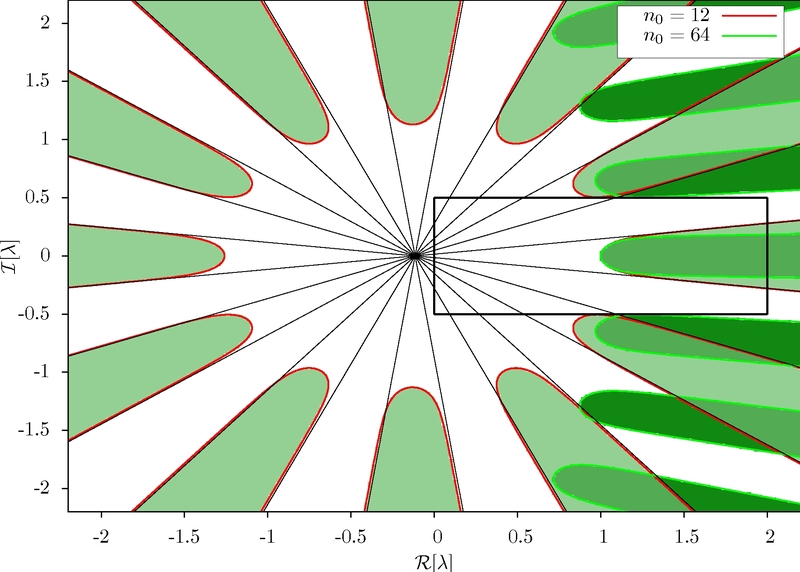}
\includegraphics[width=8.1cm]{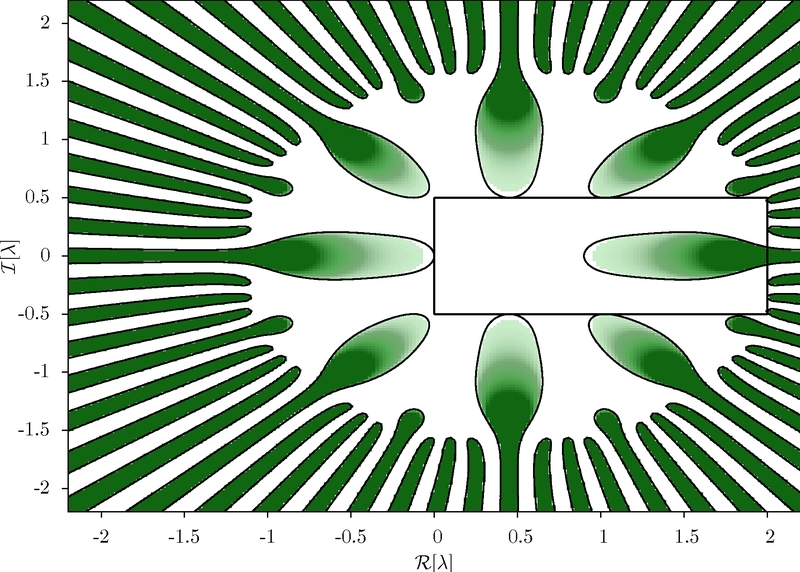}	
\caption{
       The first plot shows the region $\{\lambda | \re{p(\lambda)}\geq 1\}$ and
the surrounding contours for two power polynomials obtained in the optimization
procedure ($x_\text{min}=1$).
       The black rectangle shows the assumed region $\Omegac{D_W}$ and the gray
lines are $\theta=(2l+1)\pi/(2 n_0)$ with $l=0,1,\ldots,n_0-1$ for $n_0=12$.
       The second plot shows the result of an iterated optimization of the
peeling strategy using $n_k=2$ for $k=0,\ldots,5$ (overall order of the
polynomial $64$).
       The eigenvalues in the regions with the darkest green are computed first.
	}
	\label{fig:contourpower}
\end{center} \end{figure}

Increasing $n_0$ and adjusting $r_0$ by this minimization, one observes that the
improvement of the focusing saturates at larger $n_0$.
The contour lines become almost parallel equidistant lines for large $\rho$ 
(see first plot in Fig.~\ref{fig:contourpower}).

The optimization of the focusing can be applied in each step of the peeling
strategy: one has to choose a power $n_k$ and minimize $r_k$.
The polynomial that deviates most from the power polynomial of the same
order has $n_k=2$ for all $k$. It is shown in the second plot of 
Fig.~\ref{fig:contourpower}. 
Compared to the optimal power polynomial of the same order, the region
$\{\lambda|\re{p(\lambda)}\geq 1\}$ is narrower at the parts of the spectrum
that are computed first and slightly broader for the inner parts of the
spectrum.
Keeping the overall order of the polynomial (product of the $n_k$) fixed,
one can adjust the region inside the contour. A larger $n_0$, for example, leads
to a narrowing of the contour in the inner parts of the spectrum and a
broadening in the outer parts.
For the comparison in Section \ref{sec:comparison} we have chosen $n_k=2$ for
all $k$. This seems to be the best choice for the eigenvalues of $D_W$ in the
outer part of the spectrum. In practice it is profitable to test several
different polynomials.

In practical applications some choices of the polynomial might severely lower
the precision in the multiplication.
It might, therefore, be necessary to adapt the parameters, the normalization,
and the representation of the polynomial.
This problem occurs in particular for high orders of the polynomials.

We have calculated the eigenvalues and eigenvectors of the transformed operators
$p(D_W)$ using the restarted Arnoldi algorithm provided by the ARPACK package
\cite{ARPACK}.
The eigenvalues of $D_W$ were obtained from the eigenvectors.
Note that in several calculations we have used the even-odd preconditioned
Wilson-Dirac operator instead of $D_W$. The eigenvalues of the preconditioned
operator $\lambda^{(p)}_i$ are obtained from the eigenvalues of $D_W$ using
$\lambda^{(p)}_i=2\lambda_i-\lambda_i^2$. In the region of interest this
relation is invertible.
\section{Conclusion}
\label{sec:conclusion}
The polynomials obtained with the peeling strategy lead to an efficient
calculation of the smallest (or largest) real eigenvalues of the Wilson-Dirac
operator with the Arnoldi algorithm.
As we have shown in this work they are better adapted for the eigenvalue
distribution of this operator than simple power transformations.
The efficiency of the peeling strategy has two main reasons: it circumvents the
saturation of the focusing effect in the power transformation and the narrowing
of the computed region in the outer parts of the spectrum avoids a calculation
of regions with a large eigenvalue density close to the real axis.
Besides this better focusing effect, it provides also an acceleration of the
Arnoldi algorithm. We have presented a concrete procedure for the optimization 
of the parameters of the polynomials in Section \ref{sec:tech}. 

We have also tested Faber polynomials \cite{heuvelinearnoldifaber} for the computation of the lowest
eigenvalues.
They offer an interesting alternative with a similar performance as the peeling
polynomials in the outer parts of the spectrum.
A detailed comparison will be the subject of future work.

The procedure might be adapted for the eigenvalue distribution of other
operators with a spectrum in a connected region of the complex plane.
\begin{acknowledgements}
We thank Federico Farchioni, Istvan Montvay, Gernot M\"unster, Umut \"Ozugurel
and Urs Wenger for helpful comments and discussions.
This work was supported by the German Science Foundation (DFG) under contracts
Mu 757/13-2 and Mu 757/16-1, and by the John von Neumann Institute of Computing 
(NIC) with grants of computing time.
\end{acknowledgements}

\end{document}